# Exploring Human Factors in Spreadsheet Development

Simon Thorne, David Ball
University of Wales Institute Cardiff
Sthorne@uwic.ac.uk Dball@uwic.ac.uk

**ABSTRACT**

In this paper we consider human factors and their impact on spreadsheet development in strategic decision-making. This paper brings forward research from many disciplines both directly related to spreadsheets and a broader spectrum from psychology to industrial processing. We investigate how human factors affect a simplified development cycle and what the potential consequences are.

## 1.0 INTRODUCTION

Human factors are present in every activity that humans undertake. Human factors really describe the frailties of interaction and interface between man and the world. In this paper we focus on how these factors affect spreadsheet modellers. The choice of spreadsheets is not arbitrary, recent research (Fernandez, 2002 and Gosling, 2003) has shown that organisations rely heavily on the use of spreadsheets to make strategic decisions and that many business critical processes are implemented using spreadsheet applications. This sort of reliance on spreadsheets is tactically dangerous, considering the issues that arise from Human Factor research.

Consider a simplistic development cycle consisting of: Plan, Build and Test as shown in figure 1. Throughout this cycle there are a number of different human factors that will impact on the quality and integrity of the model developed. This paper will explore the factors that effect the development cycle.

Simplistic development cycle

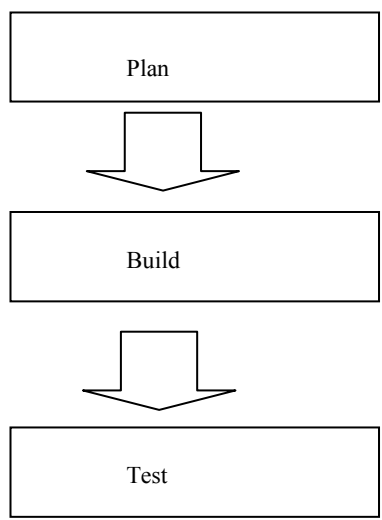

**Figure 1**

## 2.0 PLANNING STAGE

The effective planning of an information system is paramount to its success. In order to effectively plan an information system, one would have to be a trained information system professional. This leads us logically to examine the profile of the people developing spreadsheets and asses the skills they have in planning spreadsheet development.

## 2.1 SPREADSHEET DEVELOPERS

There is no typical spreadsheet developer in the modern business world. The reason for this is the great flexibility that spreadsheets offer, allowing a range of professionals to develop them. Most spreadsheet developers are end user developers by definition. End User Development (EUD) is the process of allowing end users to develop applications, using end user tools, to enhance business in some way. As end users, they will not necessarily be trained IS professionals. Indeed, many end users use computers as a necessity to perform their job satisfactorily, and EUD provides an enhancement to normal activities. EUD as an activity is paramount to information systems development and even software development. As end users, spreadsheet developers are not trained as software engineers and hence they have no knowledge of structured methodologies or processes that constitute software development. The consequences of developing software with no methodology came to a head in the 1980's with the 'Software Crisis'. There was a large upheaval of processes and standards by the industry to improve the quality of software using structured design techniques. Uptake of these methods has been widespread in the software industry. The same standards were not applied to EUD, although there was research published in the 1980's that proposed frameworks for the management of EUD (Brown and Bostrom, 1989 Munro *et al.*, 1987 and Alavi *et al.*, 1987). As was observed by Gosling (2003), there has been little uptake on these management strategies and hence spreadsheet development is ad-hoc and chaotic. Figure 2 highlights this problem. The graph is the response to the question "Do you apply a methodology when developing your spreadsheet?"

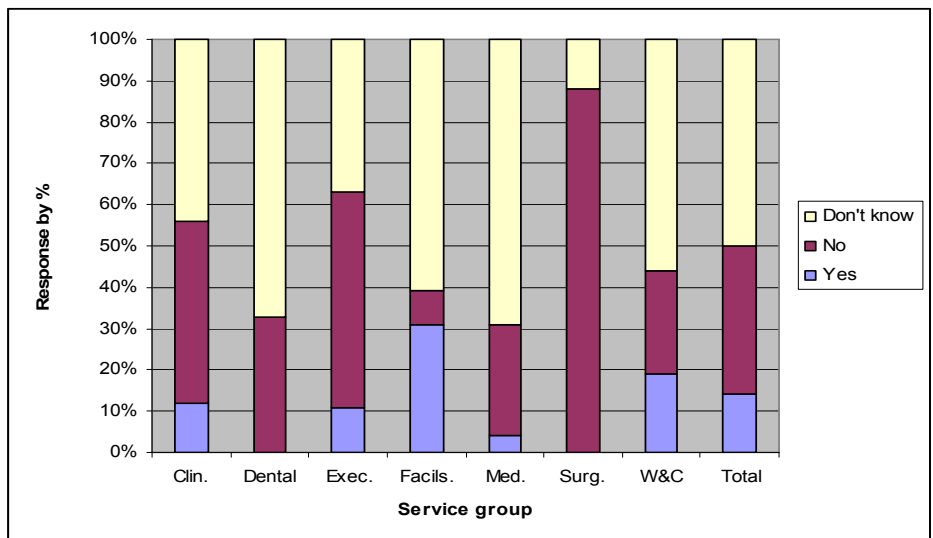

Figure 2 (Gosling, 2003)

This is further reflected by the findings of a large-scale investigation in to End User Computing in 34 UK organisations (Taylor *et al.*, 1998) This study highlighted the lack of training given to End Users in formal information systems methodological approaches. Application of such management strategies, as Alavi *et al.* propose, is proven to improve the quality of the end product; the improvement in software quality after the software crisis is strong evidence of this as observed by Yourdon (1997). The fundamental problem is that the developers of software are humans and humans make mistakes especially when they are not trained properly. In addition overconfidence plays a role. Overconfident modellers under time pressure feel no need to plan, since that would take time.

## 2.2 OVERCONFIDENCE

'Often wrong but never doubting' - Anon

Overconfidence in human activity is prolific; research shows that in relatively complex problems, humans are consistently overconfident. In particular Russo and Schoemaker (1992) examine the costs and causes of overconfidence in decision making. This overconfidence does not just apply to novice or inexperienced professionals. The same rules apply to 'experts' as Lusted (1977) and Oskamp (1965) demonstrated with physicians and clinical psychologists respectively.
Overconfidence, when seen from a spreadsheet development point view, is really concerned with all stages of the plan, build, and test development cycle. An overconfident spreadsheet developer will not plan, test or even question the validity of the work they have produced. This results in poorly designed, untested applications that are potentially full of errors. This practice is obviously risky in normal circumstances but consider more strategic applications and this practice becomes critical.

Research into spreadsheet modellers and overconfidence has shown that both individuals and groups demonstrate chronic overconfidence (Panko, 2003). Panko found that overconfidence ranged from 80% to 100% with the emphasis of overconfidence with individual modellers. Panko also extended his research to measure the effect of presenting the participants of the experiment with evidence that all spreadsheet developers were overconfident in their efforts. This resulted in a slight improvement in percentage accuracy of spreadsheet models developed and reduced the overconfidence of participants. This minor improvement after stark and blatant warning demonstrates the pervasiveness and severity of the problem. Burnett *et al.* (2003) also recognised overconfidence as a problem and demonstrated how software engineering principles could improve the accuracy and efficiency of spreadsheet developer models. Burnett's experiment applied a testing methodology to a number of participant created spreadsheets. The results showed that the participants managed to correct 92-96% of errors found i.e. they audited their work and discovered errors, which they corrected. This is clearly an effective method of catching and correcting errors. What this study does not reveal is the number of errors that the participants did not detect, which, according to reported statistics could be between 80-100%. Burnett also provided an extensive testing methodology (Burnett *et al.*, 2001) based upon software engineering techniques which improved the quality of spreadsheets by considering them in terms of executable and non-executable programs. Utilising this method, participants achieved between 0 and 31% (detected) error on a select number of problems.

The testing methodologies proposed by Burnett *et al.* are accurate and effective methods of catching errors. However, we return to the original question, if the end users are overconfident in the first place, will they consider the need for testing?

## 3.0 BUILD STAGE

At the build stage there is the greatest opportunity for human factors to impact on the quality and integrity of the spreadsheet. In this segment we explore several fundamental cognitive human issues that affect accuracy in development and even how we interact with spreadsheet programs.

## 3.1 HUMAN LEARNING AND MEMORY

As Gross (2001) observes, human learning is a hypothetical construct that cannot be observed directly but is implied in the improvement of cognitive and mechanical skills. According to Howe (1980), learning involves perception and memory. It is well accepted by the psychological community that learning and memory are essentially the same thing. The actual mechanical process of learning is based on principles of pattern matching and learning from experience, that is to say adjusting behaviour based upon previous experience to gain some more desirable result.

The 'cognitive load' (the demand of cognitive tools for a specific task) is also important. According to Kruck et al. (2003) the cognitive load is based upon four interlocking supersets: Skill Character; Working memory; Long-term memory and Task Demand. Within these supersets there are several subsets such as problem solving, memory load and accuracy. Assessing each subset in each superset allows one to build a picture of the cognitive load for a given task. Kruck et al. applied this method to a number of different everyday tasks that ranged from typing to routine medical diagnostics. He also applied this method to spreadsheet tasks, the results of this underlines the high cognitive demand on spreadsheet modellers, see table 2.

|  | Skill Character | | | Working memory | | Long term memory | | Task demands | |
|---|---|---|---|---|---|---|---|---|---|
| Tasks | Problem solving | Perceptual motor | Planning | Unit task structure | Memory load | Input to LTM | Retrieval from LTM | Pacing | Accuracy |
| Typing | Low | High | Low | Low | Low | Low | Low | Low | Int. |
| Driving a car | Low | High | Int. | Low | Int. | Low | Low | High | High |
| Mental multiplication | Int. | Low | Low | High | High | Low | Int. | Low | High |
| Balancing check book | High | Low | Int. | High | High | Int. | Int. | Low | High |
| Writing a business letter | High | Low | High | High | Int. | Int. | Int. | Low | Int. |
| CPA doing income tax | High | Low | High | High | Int. | Int. | High | Low | High |
| Routine medical diagnostics | High | Low | High | High | High | High | High | Int. | High |
| Spreadsheet tasks | High | High | High | High | High | Low | High | Low | High |

Table 2 (Kruck et al., 2003)

Kruck et al. were experimenting to determine if training spreadsheet users would affect their accuracy. Kruck concentrated on four elements of cognition that was labelled a framework for cognitive skills, see figure 3.

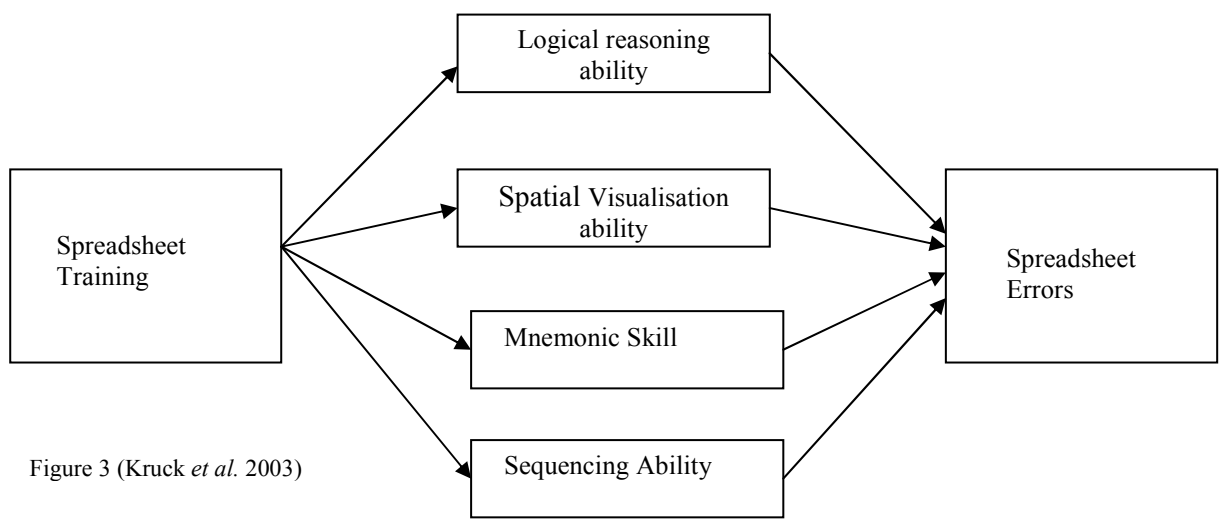

Figure 3 (Kruck et al. 2003)

It was found that the only element that improved significantly after training was logical deduction. Kruck *et al.* found that by improving the participant's logical reasoning skills, the quality of the spreadsheets they developed was higher i.e. they had fewer errors.

From a cognitive point of view, developing spreadsheets is a method of developing software using syntax and logical constructs in much the same way writing a program in C++ or Visual Basic. In the latter two, the developer will have to learn the syntax and constructs over a period of time before the program will achieve what is required. Spreadsheets were designed to be a tool that could be utilised by non-information systems professionals. This has both advantages and disadvantages, while the user can develop a spreadsheet with no formal training; the probability of novice developers making mistakes is heightened. Consider a novice user modelling a spreadsheet; the interface is intuitive and logical and the use of complicated syntax is initially limited. Once some level of problem complexity is breached, the user will have to begin using more complicated statements. In spreadsheet applications, the syntax of arguments is often complicated and the flexible intuitive interface is replaced with a single line of code for a formula. Considering the user is not a software engineer or a programmer, this interface has serious implications on how effective the user will be at manipulating the environment to best suit their needs. Research has shown that a limited natural language interface impacts on the effective programming and learn-ability of an application (Napier *et al.*, 1989). There is also a negative effect on the users programming ability, where the user has little knowledge of the commands involved in the environment (Napier *et al.*, 1992). In Particular the users will consistently make mistakes when developing spreadsheet applications beyond some level of complexity (Thorne *et al.* 2004). That is not to say only Novices make errors, indeed there is strong evidence to the contrary (Oskamp, 1965)

Human working memory is also an area that has been under investigated in EUD. The principles of 'Miller's threshold' (Miller, 1956) state that when considering a problem, the subject will start to make errors after they are manipulating greater than 9 concepts simultaneously. If we consider this in terms of spreadsheet applications, these sort of complex and abstract formulae are common. Considering the problematic syntax and the abstract nature of programming formulae in spreadsheets, Miller's concept danger limit is particularly important. Whilst it is difficult to interpret Millers use of concepts into the spreadsheet paradigm, one could view concepts on a cell-by-cell basis. Using that analogy, concepts would be elements of a formula in a cell. Considering the complex argument structure of spreadsheet applications, spreadsheet modellers must routinely breach Miller's threshold. If we apply the principles of Halstead's difficulty (Halstead, 1977) to a spreadsheet formula, the complexity becomes apparent as this method breaks down formulae into operands and operators thus providing us with the concepts used in an argument (Thorne *et al.*, 2004).

Added to Miller is the work of Michie (Michie *et al.*, 1989) who demonstrated the poor link between the pairing of human and machine strengths based upon an appreciation of human learning and cognitive processes. Michie argued that the human computer interaction was fundamentally limited due to the way in which humans interact with the computer. Michie essentially argued that the roles of machine and human in interaction did not exploit either's strengths. His points are still relevant today, as the method in which we interact has not changed significantly since the paper was written.

## 3.2 HUMAN ERROR

Human error, unlike some of the other topics in this paper, is sourced from many different disciplines. Psychology initially started the interest but since there have been many disciplines interested in this phenomena. Reason (1990) produced the 'Generic Error Modelling System' (GEMS) based upon an understanding of human error taken from many disciplines. Reason proposed that errors are made on one of three levels: Knowledge based, Rule based or Skill based. Rasmussen (1986) laid the foundations of this when investigating human error in industrial processing plants. Using these paradigms we can classify EUD error and thus target counter measures to manage EUD activities more effectively. Further, Fraser and Smith (1992) investigated the errors created when comparing human behaviour to the norms of probability, casual connection and logical deduction. This research yielded evidence of humans making mistakes in simple and repetitive tasks, known as Base Error Rate (BER). This concept states that regardless of the simplicity and repetitive nature of a task, there will always be base level of error present. This phenomenon was observed in an experiment where participants were required to match colours with colour names correctly. Reason (1990) also found evidence of this phenomena and was bought to attention in EUD by Panko (1998). There are other numerous examples that include BER in spelling and grammar, calculation tasks, prediction and interpretation. Evidence gathered by Panko (2005) demonstrates a wide range of quantities, which vary significantly depending on the task. For example a typical rate observed in spelling BER ranges from 0.5% to 2.4% (errors per word). Indeed Panko concludes that a reasonable estimate for BER in any simple activity is 0.5%. In comparison more complex tasks such as programming yield a BER of around 5%. This suggests a fairly a relationship between complexity of task and BER – the more demanding a task is, the higher the level of BER.

## 3.4 MODELS AND PARADIGMS OF HUMAN COMPUTER INTERACTION

Human Computer Interaction is a wide discipline that includes physical, biological and technical aspects of Interaction between user and computer. There are several models of interaction that exist, the most popular being execution-evaluation cycles, Norman (1988), this method breaks interaction into seven stages. These stages describe the sequential process of a user planning, implementing and evaluating their work. This model represents the process of *interaction* between the computer and user, when the user has a specific task in mind, i.e. Print out the report. It can however, be applied to EUD, since the user will follow the model to produce a spreadsheet or a database using approximately the same steps. In examining the stages, it is suggested that some of the processes are not followed in EUD. The lack or misinterpretation of certain stages, i.e. evaluating the system state in respect to the goals and intentions is one of the causes of poor quality in EUD systems. Indeed, the above statement is open to all kinds of interpretation, whether it is the bias, as described by Fraser and Smith (2003), that causes the user to incorrectly interpret results or the failure to adequately test the system due to a lack of knowledge of structured methodologies.

An alternative way of interpreting human computer interaction, even spreadsheet development, is centred on problem driven modelling. Put simply, the user utilises the computer and their own cognitive faculties to solve a problem. The spreadsheet application is essentially the implementation of the cognitive model developed beforehand by the user. Problem solving in humans can be viewed as problem space searching as Newell and Simon (1972) suggested. Problems state space searching is the process of forming a goal state (what the user wants to create) a current state (the point that the user currently resides at) and the valid operators to change the current state to the goal state. The goal state in this context could be general or specific. It could be to create a spreadsheet that represents a business problem or more specifically the sum of two cells in a spreadsheet to produce a total. Consider the latter example, the goal state is a formula that sums two cells; the current state is nil (there is no part of the formula produced). The valid operators could be mathematical symbols (+ - / *), cell names and addresses (C1, B1 etc) and the applications specific operators (SUM). In this instance the problem space allows more than one valid goal state, there are several ways of writing a formula that will sum two cells. It is now at the users discretion to decide on the goal state that they desire. Selecting the best goal state presents the user with some significant problems. How does the user decide which is the *best* solution to the problem or are they even aware that there are other valid goal states. This paradigm of problem solving is utilised in machine learning techniques since this model of problem solving lends itself to the area. When machines are presented with multiple goal states, the machine will asses according to efficiency, this may take the form of the goal state that is the most compact or requires the least processing power. For a human to asses the goal states in the same way would be problematic. In this simple example the human could make the decision but in larger problems where there may be tens or hundreds of valid goal states, it would take the user a significant amount of time to resolve to the *best* solution. It is this kind of problem that humans are weak at, evaluating large amounts of data in terms of efficiency, which contains large amounts of replication; a computer on the other hand is naturally good at this.

When we consider the way in which a user interacts with a computer, in light of problem space searching, to create an application that is a representation of a system, there are several fundamental processes. The first element is matching patterns in real–world examples and realising trends in those patterns that form some rule or judgement. The second is then manipulation of mathematics to represent that system accurately and lastly using logical deduction to classify the results accordingly. Now if we consider table 1, the natural strengths of the average human and the typical conventional computer some discrepancies arise.

|          | Pattern matching | Generating real-world examples | Manipulating mathematics | Logical deduction |
|----------|------------------|-------------------------------|--------------------------|-------------------|
| Human    | Y                | Y                             | ?                        | ?                 |
| Computer | N                | N                             | Y                        | Y                 |

Table 1

From this table we can deduce that humans are strong at generating real-world examples and pattern matching but weak at mathematical manipulation and logical deduction. Conversely, computers are strong at manipulating mathematics and logical deduction but weak at generating real-world examples and pattern matching. If we then apply this to EUD and spreadsheets in particular we can see that the current paradigm places strain on the natural weaknesses of the human and doesn't exploit the computers full potential. Table 2 shows the current paradigm in spreadsheet development.

|          | Producing formulae | Generating real world examples |
|----------|--------------------|--------------------------------|
| Human    | Weak (Current)     | Strong (Proposed)              |
| Computer | Strong             | Weak                           |

Table 2

As can be seen in table 2, the human is charged with providing the computer with the formulae, at which they are naturally weak. The computer then uses the formulae in the spreadsheet but does not exploit the massive potential that it has in terms of mathematics; it merely calculates data. A new paradigm that exploits the merits of both the user and computer would allow greater interaction. Ideally a method that would play on the strengths of both human and computer would improve the way in which the two interact. One such alternative novel solution would require the human to produce examples of attribute classifications and the machine would then deduce the function of those examples and generalise to new unseen examples. This approach has been coined 'Example Driven Modelling' (EDM) (Thorne *et al.*, 2004) which uses machine learning techniques to produce a more accurate system of creating representative systems. Machine learning, in the context of EDM, is best described as the ability to adapt and extrapolate patterns in data as defined by Russel and Norvig (2003). To be more specific, the particular branch of machine learning that interests the researcher is Neural Networks and their use in example attribute classifications of data. For example, the user provides simple examples of the problem data. This data is then fed into the learning machine and it produces an equivalent model of the problem. Thorne *et al.* (2003) discussed an experiment to test the relative levels of accuracy gained from both traditionally modelling a formulae and utilising an EDM approach, over successively more difficult written problems. The results of this study found that producing the formulae with the traditional method was error prone (80% of models with error). The results of the EDM method yielded a much lower error rate (2% of models with error). These findings are reinforced by Michie *et al.* (1989) who compared human and machine learning over a series of experiments. A learning machine and human were given information regarding the legality of Rook – King moves in Chess. Both participants were given the same information and then Human and Machine learning was determined and compared. This experimentation revealed that machine learning can be more effective and efficient than human learning; the machine was consistently more accurate making *better* use of the information made available to it. Much of Michie's work throughout the 1970's to 90's was concerned with Machine Learning Techniques (MLT) and the comparison of those techniques with equivalent human abilities (Michie, 1979 and Michie, 1990). He also revealed insights into the human learning process through his work, which he tried to represent through MLT (Michie, 1982). By summarising Michie's work, the general goal of his research was to exploit human learning concepts via symbolic artificial systems to provide some machine or method that could learn more effectively. Michie found that machine learning was highly accurate, when compared to human learning, but that it was often too specialised. The machines could only ever perform a small number of tasks satisfactorily and beyond their domain, there were useless. In contrasts humans have greater generalised skill than specialised skill, affording them 'graceful degradation' in skills and knowledge. It is perceived that a novel approach such as EDM could greatly improve accuracy by delegating much of the work to the computer rather than the user.

## 4.0 TESTING STAGE

The testing stage is the final point before the user decides that the model they have produced is adequate for the task it was designed for. In addition to lack of formal testing methodologies that is implied since most spreadsheet developers aren't IS professionals, bias is considered in testing.

Gilovitch *et al.* (2002) considers the heuristic methods and bias implicit in everyday life. The most relevant parts of this text refer to the bias present in seemingly objective judgements. For example, the trend towards predicting an outcome favourably due to the fact that the subject has a vested interest in the outcome (Armor and Taylor, 2000). Further, Fraser and Smith (1992) also discusses the concept of hypothesis fixation where once a hypothesis is formed about a decision, the subject will misinterpret the results to show that their hypothesis was correct. Fraser also examines confirmation bias, which is closely linked with hypothesis fixation. Confirmation bias is the tendency for individuals to test their own work with conditions that favour a positive response. Consider the flexibility and variable accuracy of End User Developed spreadsheet applications and the above factors become critical. The issue of bias becomes critical once the user is required to evaluate their work. Pryor (2004) theorised that End Users test their spreadsheets by using the 'sniff' test. That is to say, if the figures roughly match what they are expecting then the spreadsheet is valid. If we consider hypothesis fixation, confirmation bias (Fraser and Smith, 2002) or optimistic bias (Armor and Taylor, 2002) in this context, the little testing that is applied can be rendered invalid.

# 5.0 CONCLUSIONS

Although this paper has considered the fundamental issues in Human Factors, it is not exhaustive. There may be other equally important factors that this paper has not covered that impact on spreadsheet development.

If we re-visit the original Plan Build Test analogy (see section 1.0), we can demonstrate at which stage the particular human factors play a significant role. Figure 3 shows the modified diagram.

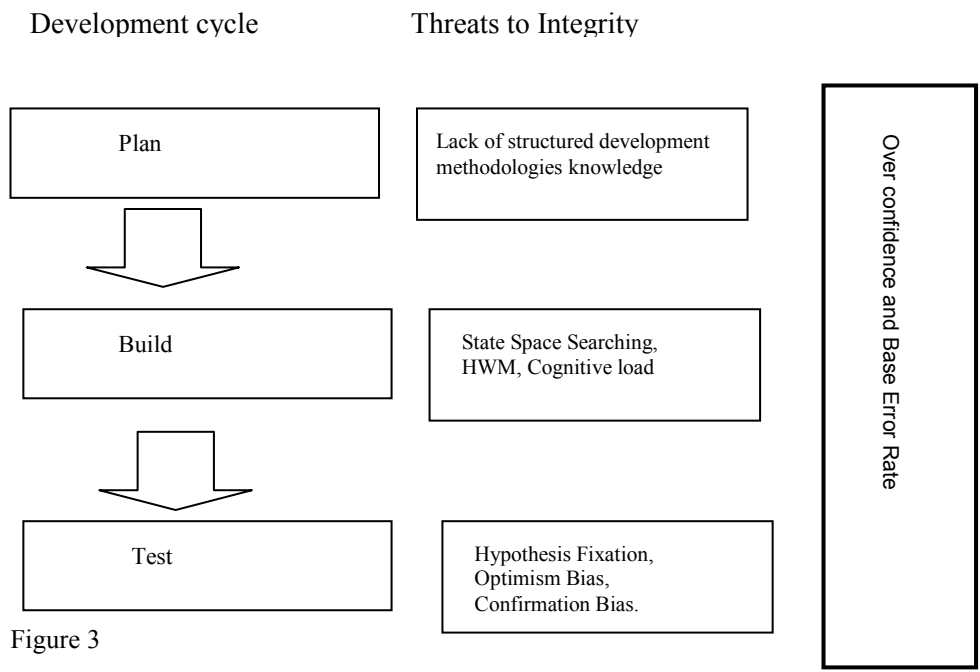

Figure 3

This diagram shows the particular issues that arise at each stage of development. For Example the greatest threat to integrity at the plan stage is the lack of structured development methods knowledge.

In addition to the stage specific threats, there are overarching issues that discretely affect each stage. The two overarching factors are Overconfidence and Base Error Rate. For example a modeller may be overconfident in planning their model, they may spend a minimal amount of time constructing a plan if at all. The same applies to testing; they may test their model inadequately due to the fact that they are confident that their model is *accurate*, indeed the plan stage may well be dismissed altogether if the modeller is overconfident. Base Error Rate (BER) plays a similar role, a user will be pre disposed to BER whilst they are building the spreadsheet but also when testing it.

In conclusion, Human Factors play a significant role throughout the spreadsheet development cycle. Further, Human Factors should be investigated further to allow the spreadsheet-modelling world to build applications and processes that are sympathetic to such issues.